# Temperature gradient-driven magnetic skyrmion motion

Eleonora Raimondo[1], Elias Saugar[2], Joseph Barker[3], Davi Rodrigues[4], Anna Giordano,[5] Mario Carpentieri[4], Wanjun Jiang[6,7], Oksana Chubykalo-Fesenko[2], Riccardo Tomasello[4*], Giovanni Finocchio[1*]

[1] Department of Mathematical and Computer Sciences, Physical Sciences and Earth Sciences, University of Messina, I-98166, Messina, Italy

[2] Instituto de Ciencia de Materiales de Madrid, CSIC, Cantoblanco, 28049 Madrid, Spain

[3] School of Physics and Astronomy, University of Leeds, Leeds LS2 9JT, United Kingdom

[4] Department of Electrical and Information Engineering, Politecnico of Bari, 70125 Bari, Italy

[5] Department of Engineering, University of Messina, I-98166, Messina, Italy

[6] State Key Laboratory of Low-Dimensional Quantum Physics and Department of Physics, Tsinghua University, Beijing 100084, China

[7] Frontier Science Center for Quantum Information, Tsinghua University, Beijing 100084, China

*corresponding authors: riccardo.tomasello@poliba.it, gfinocchio@unime.it

**Abstract**

The static and dynamic properties of skyrmions have recently received increased attention due to the potential application of skyrmions as information carriers and for unconventional computing. While the current-driven dynamics has been explored deeply, both theoretically and experimentally, the theory of temperature gradient-induced dynamics - Skyrmion-Caloritronics - is still at its early stages of development. Here, we move the topic forward by identifying the role of entropic torques due to the temperature dependence of magnetic parameters. Our results show that, skyrmions move towards higher temperatures in single-layer ferromagnets with interfacial Dzyaloshinski-Moriya interactions, whereas, in multilayers, they move to lower temperatures. We analytically and numerically demonstrate that the opposite behaviors are due to different scaling relations of the material parameters as well as a non-negligible magnetostatic field gradient in multilayers.

We also find a spatially dependent skyrmion Hall angle in multilayers hosting hybrid skyrmions due to variations of the thickness dependent chirality as the skyrmion moves along the temperature gradient.

I. INTRODUCTION

Magnetic skyrmions have been receiving increasing attention in the last decades. Since the first theoretical studies and predictions on the fundamental and promising application properties of these "topologically protected" solitons [1–3], great efforts have been placed in material development for the experimental stabilization of skyrmions. The first experimental observations were in $B_{20}$ compounds, where the bulk Dzyaloshinskii-Moriya interaction (DMI) promotes the formation of Bloch skyrmions [4–9]. Later, an increased effort has been devoted to thin films and heterostructures where ferromagnetic layers (FM) are coupled with heavy metals (HM) which are characterized by interfacial DMI (IDMI) [10]. These systems allow skyrmions to be stabilized at room temperature, which is a fundamental requirement for practical applications. Lately, skyrmions have been reported in a variety of thin film materials, including HM/single-layer FM/oxide [11–13], HM1/FM/HM2 multilayers [14–21], HM/ferrimagnet/oxide multilayers [22,23], synthetic antiferromagnets (SAFs) [24–26], as well as bulk systems [27,28].

The standard approach to manipulate skyrmions is by electrical currents through the conventional spin-orbit torques (SOT) [13,15,16,29–31]. However, alternative methods have been proposed, such as by external field [32,33] and perpendicular anisotropy [34] gradients as well as thermal gradients [35,36]. The manipulation by thermal gradients is particularly promising due to its low energy consumption, and is at the basis of a new research direction named Skyrmion-Caloritronics [35].

Thermal fluctuations have also been predicted to induce internal deformations, breathing excitations, and Brownian diffusion [37–41]. The latter has recently been experimentally observed and exploited in a low pinning material to design a reshuffle device for probabilistic computing [42]. Moreover, experiments have shown unidirectional diffusion of skyrmions in thermal gradients, where they moved from hot to cold regions [35]. This observation has been explained through the combination of repulsive forces between skyrmions, thermal SOTs [43], magnonic spin torques [44] as well as entropic forces. Recently, Gong et al. [45] proposed the existence of spin-currents generated by thermal gradients to explain the observed skyrmion motion and concluded that skyrmions can move in either direction of the temperature gradient depending on the material parameters.

We have previously demonstrated the influence of thermal fluctuations on skyrmion stationary properties in single layers [40], through variations of the magnetic parameters (exchange, interfacial DMI, perpendicular anisotropy constants) whose scaling relations were obtained by atomistic calculations. However, that study did not consider the influence of thermal fluctuations on the skyrmion dynamical properties. Moreover, in a previous study [46] performed numerically and within the Thiele's formalism [47], we have shown that a linear gradient of the perpendicular anisotropy generates skyrmion motion in a single-layer FM. We observed that the skyrmion exhibits a Hall effect with a major component of the velocity along the direction perpendicular to the anisotropy gradient, and a smaller velocity component (damping-dependent) in the same direction of the current. Furthermore, the adiabatic adjustment of the skyrmion size in the anisotropy gradient produces a uniform acceleration.

In this work, we study the effects of thermal gradients on skyrmion motion in different systems (single-layer FM with IDMI, multilayer, and SAF) inspired by the experimental results of Ref. [35] and from a fundamental point of view. We consider the previously obtained 2D scaling relations to analyze the thermal gradient-driven skyrmion motion in a single-layer with IDMI. Using atomistic simulations, we also compute the characteristic scaling relations for magnetic multilayers where several atomic layers are taken into account and the DMI is induced at both ferromagnet interfaces (for example Pt/FeCo/Ir [17,48] can be used as the active trilayer in *N* repetitions for a multilayer stack) and investigate the thermal gradient-driven motion of a skyrmion in these systems. In particular, we focus on two effects: the entropic torque and magnonic torque. The first one, already studied for domain walls [49], is analyzed in a deterministic scenario and is based on thermodynamic phenomena. It generates a skyrmion motion towards the region where its free energy is minimized. The second one occurs in a stochastic framework where thermal spin-waves propagating from the hot to the cold region exert a torque on skyrmions by moving them in the opposite direction – from cold to hot [44]. We observe that, when driven by the entropic torque, skyrmions in single layers move from the cold to the hot region with a finite skyrmion Hall angle, while in multilayers they move in the opposite direction (from hot to cold region). The numerical results are corroborated by a generalized Thiele's equation developed for this scenario, taking into account variations of the skyrmion size along its trajectory. Moreover, we show that the skyrmion Hall angle is completely suppressed in SAFs, similarly to the current-driven skyrmions [50,51]. The effect of the magnonic torque agrees with previous predictions [44] and induces the skyrmion motion towards the hotter region regardless the system.

Our results have fundamental implications in the future development of skyrmionic devices combining thermal gradients and SOTs where the proper temperature dependence of the parameters should be taken into account. In addition, our work extends the application of controlled skyrmion motion to electric insulators and contributes to the design of waste heat recovery methods taking advantage of the thermal-driven skyrmion nucleation and shifting, promoting the field of Skyrmion-Caloritronics.

## II. THEORETICAL MODELS

### A. Micromagnetic simulations

The micromagnetic computations were performed with a state-of-the-art micromagnetic solver, PETASPIN, which uses the Adams-Bashforth [52,53] method to numerically integrate the Landau-Lifshitz-Gilbert (LLG) equation:

$$\frac{d\mathbf{m}}{d\tau} = -(\mathbf{m} \times \mathbf{h}_{\text{eff}}) + \alpha_G \left( \mathbf{m} \times \frac{d\mathbf{m}}{d\tau} \right) \quad (1)$$

where $\alpha_G$ is the Gilbert damping, $\mathbf{m} = \mathbf{M}/M_s$ is the normalized magnetization vector, and $\tau = \gamma_0 M_s t$ is the dimensionless time, with $\gamma_0$ being the gyromagnetic ratio, and $M_s$ the saturation magnetization. $\mathbf{h}_{\text{eff}}$ is the effective magnetic field, which includes the exchange ($A$), interfacial DMI (IDMI $D$), uniaxial anisotropy ($K_u$), magnetostatic, and external fields ($H_{ext}$) [21,54]. We include the change in saturation magnetization with temperature as $M_S(T) = M_S(0)\left(1 - \left(\frac{T}{T_{\text{lim}}}\right)^\delta\right)$ with $M_S(0)$ being the saturation magnetization of the ferromagnet at zero temperature, $\delta = 1.5$ and $T_{\text{lim}} = 1120$ K is the Curie temperature [55–57].

We simulated three different systems: single-layer FM with IDMI, multilayer with 5 FM repetitions, and SAF. The parameters at zero temperature used for each of them are listed in Table I, while further details of the micromagnetic model can be found in the Supplemental Note 1 [58].

Table I. Summary of the micromagnetic parameters at zero temperature used in the simulations.

| Parameter | Single-layer FM | Multilayer | SAF |
|---|---|---|---|
| $M_s$ (kA/m) | 1060 | 1300 | 770 |
| $A$ (pJ/m) | 20 | 15 | 20 |
| $D$ (mJ/m$^2$) | 2.2 | 1.0 | 2.5 |
| $K_u$ (MJ/m$^3$) | 0.90 | 1.20 | 0.60 |
| Out-of-plane $H_{ext}$ (mT) | 15 | 60 | 0 |

For the macroscopic parameters, in a single-layer FM with IDMI, we used the 2D scaling relations from our previous work [40] which are as follows:

$$A(T) = A(0)m(T)^\alpha, \quad D(T) = D(0)m(T)^\beta, \quad K_u(T) = K_u(0)m(T)^\gamma, \quad (2)$$

where $\alpha = \beta = 1.5$, $\gamma = 3.0$, and $m(T) = M_S(T)/M_S(0)$. For the Pt/FeCo/Ir trilayer, the scaling relations were evaluated by atomistic simulations, as shown below.

In the micromagnetic simulations, we used a linear thermal gradient from 100 K to 300 K along the *x*-direction. The minimum (maximum) value is considered on the left (right) side of the sample, where the corresponding maximum (minimum) value of the magnetic parameters are calculated from the scaling relations.

B.    Atomistic Simulations

We used atomistic spin simulations to calculate the temperature dependence of magnetic parameters using the technique developed in Ref. [40]. We model the thin film as an FCC (111) lattice with 2 layers of Fe (adjacent to Ir) 1 layer of 50:50 mixed Fe and Co and then 2 layers of Co (adjacent to Pt). The Hamiltonian is given by

$$H = \sum_{<ij>} J_{ij} \mathbf{S}_i \cdot \mathbf{S}_j + \sum_{<ij>} \mathbf{D}_{ij} \cdot (\mathbf{S}_i \times \mathbf{S}_j) - \sum_{<ij>} K_{ij} S_{zi} S_{zj} \quad (3)$$

where the first term is the isotropic exchange, the second is the DMI which only occurs at the Ir-Fe and Co-Pt interfaces and the third is the two-ion anisotropy which acts only at the Co-Pt interface. In this finite thickness film, the IDMI (and anisotropy) is only present at the Ir-Fe and Co-Pt interface layers whereas the exchange is present in and between all layers. The angle brackets, <ij>, denote that we only consider nearest neighbors. We parametrize the exchange anisotropy and DMI constants using values from ab initio calculations in Ref. [17] $J_{ij}(Fe-Fe) = 24.2\,\text{meV}$, $J_{ij}(Co-Co) = 29.0\,\text{meV}$, $J_{ij}(Fe-Co) = 26.6\,\text{meV}$, $D_{ij}(Fe-Ir) = -0.854\,\text{meV}$, $D_{ij}(Co-Pt) = 1.281\,\text{meV}$, $K_{ij}(Fe-Fe) = 0.59\,\text{meV}$ and used magnetic moments $\mu(Fe) = 1.7\mu_B$ and $\mu(Co) = 1.3\mu_B$ [59].

The temperature dependence of the anisotropy was calculated using constrained Monte Carlo method [60]. The temperature dependence of exchange stiffness and IDMI was calculated from the softening [61] and shifting asymmetry of the spin wave spectrum. The spectrum contains five bands and we fitted the lowest band in the low k-limit (see supplementary information of Ref. [40] for details of the technique).

We express the scaling relations in terms of the magnetization of the whole thin film, which corresponds to what is most accessible experimentally and is used in micromagnetic simulations for single layers. We found that the Ir/FeCo/Pt trilayer has the scaling parameters $\alpha = 1.7$, $\beta = 2.0$ and $\gamma = 2.5$. This is different from in a 2D (atomic monolayer) system where the α and β parameters are found to be identical [40,62]. Thus, $\beta$ and $\gamma$ are modified because their temperature dependence is dominated by spin fluctuations at interfaces which are different from the bulk where there is a different number of neighboring spins. This implies that, by changing the properties of the multilayer, including heavy metal materials and geometry, it is possible to tune the exponents $\beta$ and $\gamma$. We note, however, that $\alpha$ is closer to the bulk value $(\alpha = 1.8)$ [63] than in the 2D case $(\alpha = 1.5)$ due to the finite thickness of the model considered. The exchange and DMI are also sensitive to different types of spin fluctuation, with $A$ being mostly dependent on in-plane fluctuations whilst $D$ is sensitive also to the out-of-plane fluctuations which are different in the thicker and asymmetric system compared to the 2D case [62]. We emphasize, therefore, that the thickness of the film and the interface properties play an important role overall on the temperature dependence of the material parameters.

## C. Generalized Thiele's equation

Despite the complex dynamics of the LLG Eq. (1), magnetic skyrmions are collective excitation states which behave as particle-like objects with reduced degrees of freedom. This physical behavior is captured by the Thiele's equation [47,64,65] which describes the motion of rigid magnetic textures. By considering only the lowest energy excitations of the skyrmion, we assume that the magnetization dynamics is reduced to a rigid translation and change in size of the magnetic skyrmion,

$$\frac{d\mathbf{m}}{d\tau}(\mathbf{x},\tau) = -\left(\mathbf{v}(\tau) + \frac{d\mathbf{r}(\tau)}{d\tau}\right) \cdot \nabla \mathbf{m}(\mathbf{x},\tau) \qquad (4)$$

where we considered the skyrmion as the center of the coordinate system such that $\mathbf{r}(\tau)$ represents the position of the skyrmion boundary (i.e. where $m_z = 0$), and $\mathbf{v} = v_x\hat{\mathbf{x}} + v_y\hat{\mathbf{y}}$ is the velocity of the skyrmion center. By substituting Eq. (3) into Eq. (1) and projecting along $\mathbf{m} \times \partial_i \mathbf{m}$, with $i = x, y$, we obtain a generalized Thiele equation from which we derive the velocity of the skyrmion as

$$v_x = -\frac{4\pi\alpha_G D_{ex} F_x}{\alpha_G^2 D_{ex}^2 + (4\pi)^2} \text{ and } v_y = -\frac{4\pi F_x}{\alpha_G^2 D_{ex}^2 + (4\pi)^2}, \qquad (5)$$

where $D_{ex} = \int d^2x (\partial_x \mathbf{m} \cdot \partial_x \mathbf{m}) = \int d^2x (\partial_y \mathbf{m} \cdot \partial_y \mathbf{m})$ is the viscosity tensor components for a radially symmetric skyrmion, and $F_x = -\partial_x V$, with $V(\mathbf{x}') = \int d^2x (\mathbf{m}(\mathbf{x}',\tau) \cdot \mathbf{h}_{\text{eff}}(\mathbf{x},\tau))$, is the force due to a gradient in the effective field along the *x*-axis. Due to the radial symmetry of the skyrmion, variations of the skyrmion radius, i.e. $r(\tau)$, does not enter explicitly the equation for the velocity. The size of the skyrmion, however, influences $D_{ex}$ and $F_x$. The non-vanishing component $v_y$ of the velocity perpendicular to the force $F_x$ is due to the Magnus force experienced by the skyrmion and is associated with its topological charge [66]. Moreover, the velocity $v_x$ along $F_x$ is due to the influence of damping $\alpha_G$.

In the presence of a thermal gradient (expressed as the gradient of magnetic parameters) along the $x$-direction, we obtain a dependence of the effective field $\mathbf{h}_{\text{eff}}$ on the $x$-coordinate which in turn produces a motion of the skyrmion. For sufficiently large skyrmions, assuming the following ansatz $\tan\left[\Theta_0(r)/2\right] = (r_{sk}/r) e^{\xi(r_{sk}-r)}$ [40], we have that

$$V = 2\pi \left( 4A \left( \frac{\Delta}{R} + \frac{R}{\Delta} \right) + 4K_{\text{eff}} R\Delta - 2\pi DR + M_S H_{\text{ext}} R^2 \right) \qquad (6)$$

where $K_{\text{eff}} = K_u - \frac{\mu_0}{2} M_S^2$ is the effective anisotropy, $\Delta = \sqrt{A/K_{\text{eff}}}$ is the domain wall width and $R = \Delta / \sqrt{2\left(1 - \pi D/\left(4\sqrt{AK_{\text{eff}}}\right)\right)}$ is the equilibrium radius of the skyrmion [40,64,67,68]. Notice that, for a magnetization saturation varying along the sample, the magnetostatic cannot be fully incorporated into the renormalization via the effective anisotropy $K_{\text{eff}}$, such that we also included a contribution to the external field that is proportional to $M_s$, for details see Supplemental Material. Thus, a gradient in temperature produces position-dependent material parameters $A$, $D$, $K_u$ and $M_S$, as well as, within an adiabatic approximation, a position-dependent size of the skyrmion and generates the motion of the skyrmion with a non-constant velocity. Furthermore, we emphasize that the change in the saturation magnetization $M_s$ along the film due to the temperature gradient also produces a stray field gradient. This effect can be incorporated into the effective anisotropy (thin-film approximation) for the case of the single layer, however the thin-film approximation does not hold for the multilayer. Therefore, in the Thiele's equation for the multilayer, the non-negligible stray field gradient is considered through a perpendicular external field gradient.

III.  RESULTS

A.  Single-layer FM and multilayer

Figure 1 shows the first main result of this work which is obtained from deterministic micromagnetic simulations when a linear gradient of temperature is considered for the single-layer FM and the multilayer. Surprisingly, the trajectory of the skyrmion (here we consider the position of the skyrmion given by the $x$-$y$ position of the skyrmion core, calculated as the geometrical center of the circular region enclosed within $m_z \leq 0$), is qualitatively opposite for the two cases. In the single-layer FM (Fig. 1(a)), the skyrmion moves with positive velocities in x- and y-directions, i.e. towards the region with higher temperature. Whereas, in the multilayer case (Fig. 1(b)), it has negative velocities, with respect to x- and y-directions, i.e. it moves towards the colder region. The latter is a particularly interesting outcome due to the qualitative agreement with the experimental observations of skyrmion motion driven by linear thermal gradients in magnetic multilayers (see Fig. 3a in Ref. [35]).

To understand the origin of such a different behavior, we performed a detailed analysis combining full numerical simulations and Thiele's equation.

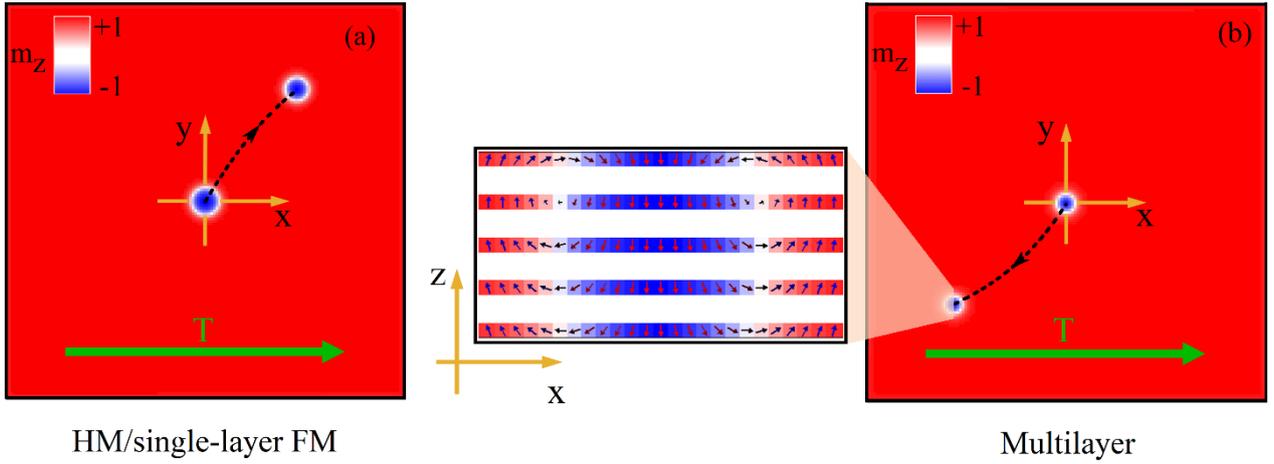

FIG. 1 Trajectory of the skyrmion induced by a linear gradient of temperature in a (a) single FM with IDMI, and (b) multilayer with 5 FM repetitions. The magnification of the skyrmion in (b) shows the cross section along the multilayer thickness that highlights the existence of a hybrid skyrmion [21].

When considering a deterministic approach with linear gradient of the micromagnetic parameters, the skyrmion moves towards the parameters region where its energy is minimized. In particular, in this work, the energy minimization as a function of temperature leads the skyrmion radius to expand. Therefore, the effect that each micromagnetic parameter gradient has on the skyrmion motion can be straightforwardly deduced since in this case it will move towards the region where the local value of the parameter favors the existence of a larger skyrmion. To confirm the previous qualitative argument, we performed systematic simulations in which the gradients of the parameters were considered independently both in the single-layer FM and multilayer. The results are summarized in Table II, while the different trajectories are shown in the Supplemental Note 2 [58]. The perpendicular anisotropy and exchange parameters gradients promote the skyrmion motion to the region where those parameters are smaller, therefore from the cold to the hot region. On the contrary, the $M_S$ (magnetostatic field), $D$ and the combination of $M_S$ and $A$ gradients, move the skyrmion to the region where those parameters are larger, thus from the hot to the cold region.

Table II. Summary of the trajectories of the skyrmion motion under a linear gradient of the parameters.

| Parameter gradient | Single-layer FM | Multilayer |
|---|---|---|
| $M_s$ | $v_x<0$, $v_y<0$ (motion to the cold region) | $v_x<0$, $v_y<0$ (motion to the cold region) |
| $A$ | $v_x>0$, $v_y>0$ (motion to the hot region) | $v_x>0$, $v_y>0$ (motion to the hot region) |
| $D$ | $v_x<0$, $v_y<0$ (motion to the cold region) | See discussion of Fig. 3 below |
| $K_u$ | $v_x>0$, $v_y>0$ (motion to the hot region) | $v_x>0$, $v_y>0$ (motion to the hot region) |

When considering the combinations of all parameters gradient due to a thermal gradient, the scenario becomes complex and the final effect on skyrmion motion depends on their values. Figure 1 suggests that the direction of skyrmion motion results from which of the antagonistic effects ($A$ and $K_u$, or IDMI and $M_S$) is dominant.

First, we focus on the single-layer FM. Figure 1(a) entails that the 2D scaling relations lead to the dominance of $A$ and $K_u$ parameters over IDMI and $M_S$. To better understand and control this behavior, we must consider two degrees of freedom: (i) the change of the scaling exponents, and (ii) different zero temperature values of the magnetic parameters. We focus on the IDMI scaling relation since the value of the IDMI can be easily tuned by modifying the materials, i.e. using different heavy metals and modifying their thickness [69], at the ferromagnetic interfaces (see section II.B) without a significant change of other material parameters. By fixing the scaled values of the other parameters as well as the zero temperature value of the IDMI, we carry out a systematic study of the skyrmion trajectory as a function of the scaling exponent $\beta$ (the range of values for $\beta$ were chosen such that the radially symmetric skyrmions were still stable). Figure 2(a) shows the results where, for $1.5 \leq \beta \leq 2.7$, the skyrmions preserves its motion towards the hot region, similarly to Fig. 1(a), whereas, in the range $2.8 \leq \beta \leq 4.0$, the motion is reversed and the skyrmion moves towards the cold region. In other words, there exists a threshold value of the exponent $\beta$ which determines the motion of the skyrmion to the cold region, albeit the DMI should decay with magnetization quite fast.

Figure 2(b) illustrates the skyrmion motion when we fixed two values of $\beta = 4.0$ and $1.5$, respectively for the left and right side of the figure, together with the scaled values of the other parameters and changed the zero-temperature value of the IDMI, $D(0)$. In both cases, the skyrmion trajectory is barely affected by $D(0)$. Therefore, we conclude that a strategy to control the direction of the skyrmion motion is to properly tune the scaling exponent of the IDMI parameter (see section II.B).

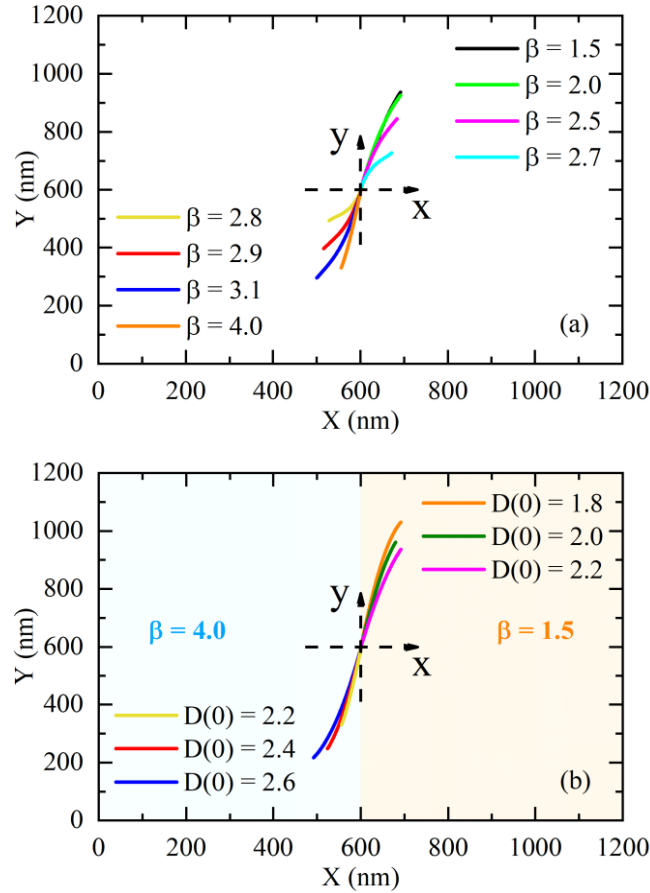

FIG. 2: (a) Trajectories of the skyrmion as a function of different $\beta$. (b) Trajectories of the skyrmion as a function of different $D(0)$ for $\beta = 1.5$ and $\beta = 4.0$, respectively.

We also performed a systematic study on the effect of the magnitude of the thermal gradient on the skyrmion velocity, which increases with the gradient as expected (see Supplemental Note 3 [58]).

Once the behavior in the single layer is fully understood, we proceed with the magnetic multilayer to understand the source of the skyrmion motion towards the colder region. In magnetic multilayers, skyrmions are characterized by an additional degree of freedom linked to the thickness-dependent chirality [21,48,70] which can be modified by tuning the IDMI parameter. It is well-documented that an increase of the IDMI parameter on the hybrid skyrmion profile shifts the central

Bloch skyrmion towards one of the most external layers, according to the IDMI sign [21,70]. Hence, we expect that, when considering a linear variation of *D*, the hybrid skyrmion profile changes along the gradient. It is important to investigate whether this effect influences the skyrmion trajectory. Thus, we fixed the IDMI gradient from 0 to – 2 mJ/m$^2$, such that in the center of the sample – corresponding to the *x-y* plane – the IDMI is given by $D = -1.0\,\text{mJ/m}^2$. In this case, we consider an IDMI gradient. At zero IDMI, static simulations show a symmetric hybrid skyrmion (see Supplemental Note 1 [58]), where the Bloch skyrmion is exactly in the middle layer.

Figure 3(a) depicts the hybrid skyrmion trajectory due to the IDMI gradient. We notice a change of slope when the *x*-position is around 540 nm. To understand the origin of such a change, we compute the helicity of the skyrmion, i.e. the in-plane angle of the magnetization vector at the boundary of the skyrmion (region where $m_z = 0$) as illustrated in the inset of Fig. 3(b). An angle of 90° corresponds to a Bloch skyrmion, while an angle of 0° (180°) corresponds to a Néel skyrmion. The initial hybrid skyrmion is placed at the center of the sample in the x-y plane, where we expect the finite value of *D* to promote the shift of the Bloch skyrmion downward. This is evident from the cross-section S1. Therefore, we calculate the angle for the skyrmion placed in the 2$^{nd}$ layer of the 5-repeats multilayer. The initial angle is about 107° in Fig. 2(b), larger than 90° still due to the finite value of *D* which tilts the Bloch chirality towards a Néel inward chirality. As the skyrmion moves towards larger values of |*D*|, the angle tends to approach 180°, which implies that the Bloch skyrmion has been shifted more downward being replaced by a Néel skyrmion with inward chirality. Our outcomes are also confirmed by the cross-sectional views S2-S5 in Fig. 3. By comparing the change of the slope in Fig. 3(a) with the corresponding angle in Fig. 3(b), we can conclude that the variation of the trajectory originates from the change of the thickness-dependent chirality of the hybrid skyrmion. This finding represents the second main result of this work.

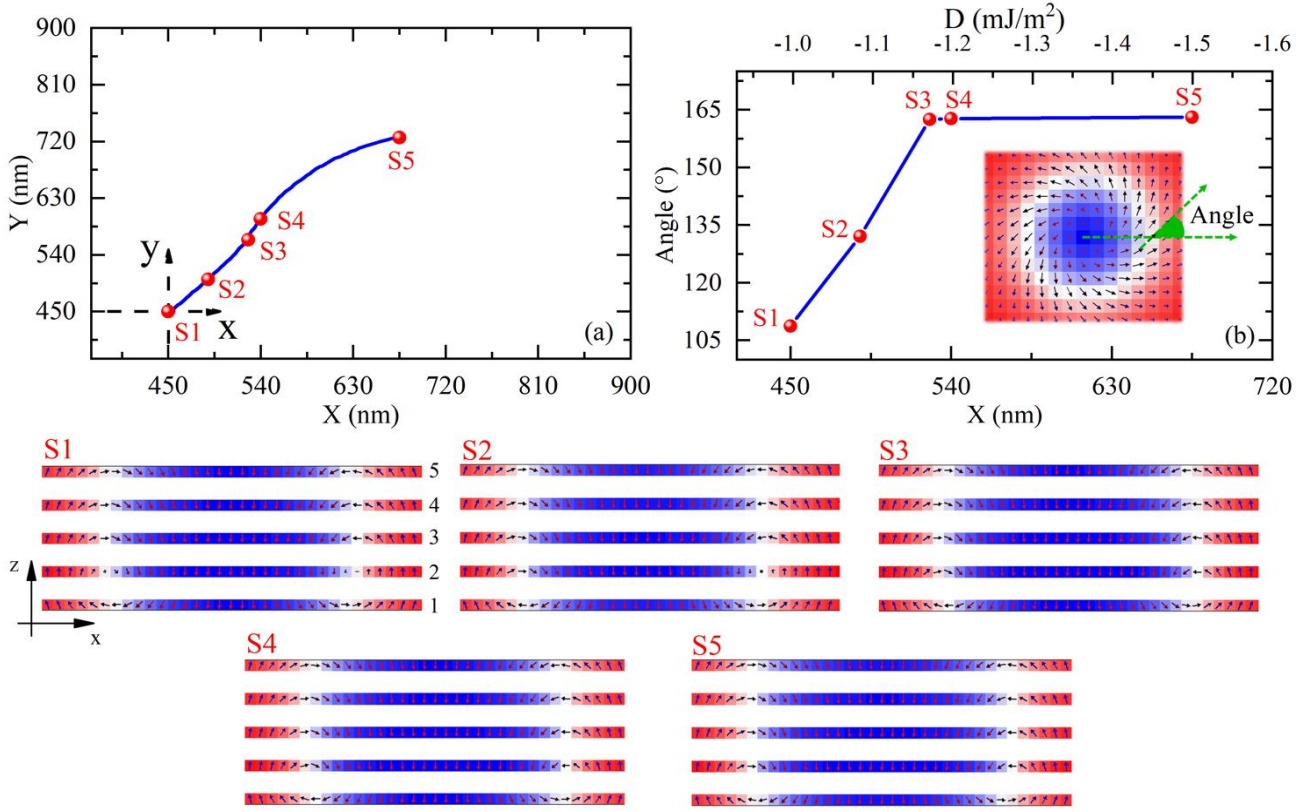

FIG. 3: (a) Skyrmion trajectory under a linear gradient of IDMI. (b) Angle of the magnetization vector of the skyrmion placed in the 2$^{nd}$ layer of the 5-repeats multilayer as a function of the *x*-position of the skyrmion core. Inset: example of skyrmions where the angle of the magnetization vector is indicated. The S1-S5 points in (a) and (b) are linked to the cross-sectional views below.

However, the change of the skyrmion profile and helicity is not enough to justify the different behaviors observed between the single-layer FM and the multilayer. Indeed, to exclude these effects, we performed a simulation fixing the thickness-dependent chirality during the motion, and still observed the skyrmion motion in the multilayer system towards the colder region.

An important difference from the single-layer FM is the value of the scaling exponents. To check their effects, we performed a single-layer simulation, similar to the one in Fig. 1(a), and considered the scaling relations obtained for the multilayer system (see section II.B). In this simulation, we observed that the Néel skyrmion moves towards the colder region (Fig. 4(a)), similarly to the hybrid skyrmion in multilayers. This unequivocally confirms the crucial role of the scaling relations that, for multilayers, leads to the dominance of the effect of the IDMI over the other parameters.

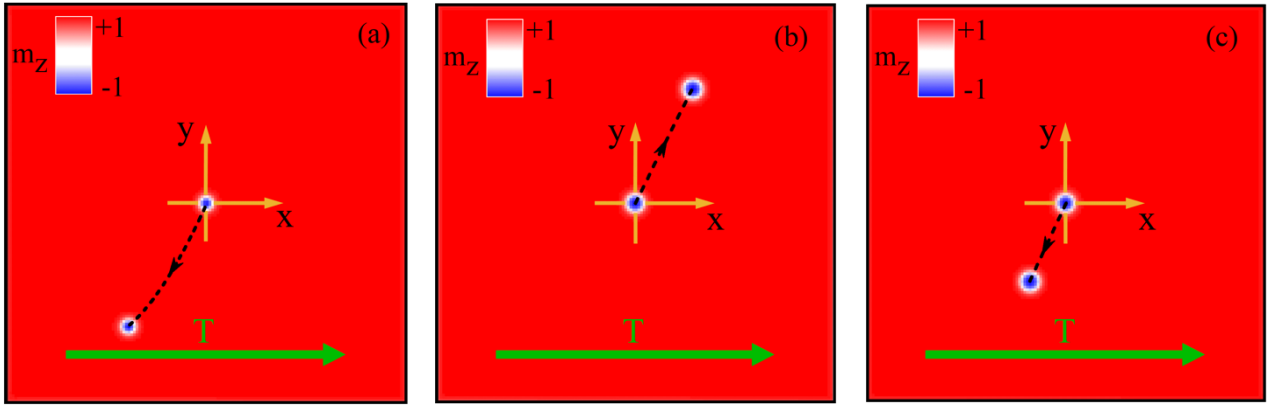

FIG. 4: Skyrmion trajectory under thermal gradient calculated for (a) a single FM layer with the scaling relations of the multilayer via micromagnetic simulations, and via the Thiele's equation with the scaling relations of the (b) single-layer FM, and (c) multilayer.

To corroborate our analysis in the single-layer FM simulations, we verified the behavior by using the Thiele's equation (see Section II.C). In particular, we considered the change of the skyrmion size with the material parameters gradients, and a $M_S$-dependent external field gradient that mimics the contribution of the magnetostatic field gradient. Figure 4(b) and (c) show the skyrmion trajectory within the Thiele formalism for the single-layer FM and multilayer systems, respectively, where the skyrmion behavior is in excellent agreement with the full micromagnetic simulations. Notice that, the skyrmion trajectory for the single-layer scaling relations is not strongly affected by the presence of the external field gradient, contrarily to the trajectory on the multilayer. In fact, we verified that removing the contribution of the stray field gradient: the skyrmion in the multilayer also moves to the hotter region, likewise the case of the single-layer FM. This outcome points out that the gradient of the magnetostatic field in multilayers has a crucial effect in determining the skyrmion motion towards the colder region.

Once the influence of the entropic torques were well-understood, we analyzed the effect of the magnonic torque. We performed stochastic micromagnetic simulations (see Supplemental Note 4 [58]) with a linear thermal field [71,72] gradient from 0 to 100 K. The skyrmion moves towards the hot region, similarly to Fig. 1(a) and to previous reports [44], but with a stochastic dispersion of the Hall angle around its average value due to thermal fluctuations. These results are similar even for the case when magnonic torque and entropic torque (scaling relations of the parameters) are considered simultaneously.

### A. SAF

In a SAF, two skyrmions of opposite topological charges are antiferromagnetically coupled via the interlayer exchange coupling [51,73–75], and a vanishing skyrmion Hall effect is achieved [50,51,76]. Therefore, we expect the linear gradient-driven skyrmion motion to occur only in the direction of the gradients (*x*-axis). This can be explained by the fact that in this case, the two coupled skyrmions experience Magnus forces of same magnitude in opposite directions, such that the total net force perpendicular to perturbation vanishes, as in the case of spin-current-driven antiferromagnetic Néel skyrmions [50].

Our calculations show that the results obtained above on ferromagnetic systems are still valid for SAFs, where, according to the material choice, either single layer or multilayer scaling relations can be used. In the following, we show only the results with scaling relation of the multilayer.

Figure 5(a) and (b) show the motion direction due to the gradients IDMI, $M_S$, or thermal gradients, while Figs. 5(c) and (d) show the opposite direction of motion due to $K_u$ or $A$ gradients. In all the cases, the skyrmion is characterized by a zero Hall angle, but the entropic torque due to the thermal gradients (Fig. 5(a) and (b)) promotes the motion towards the cold region. Here, notice that the magnetostatic field gradient does not play any role because the SAF system has no stray field.

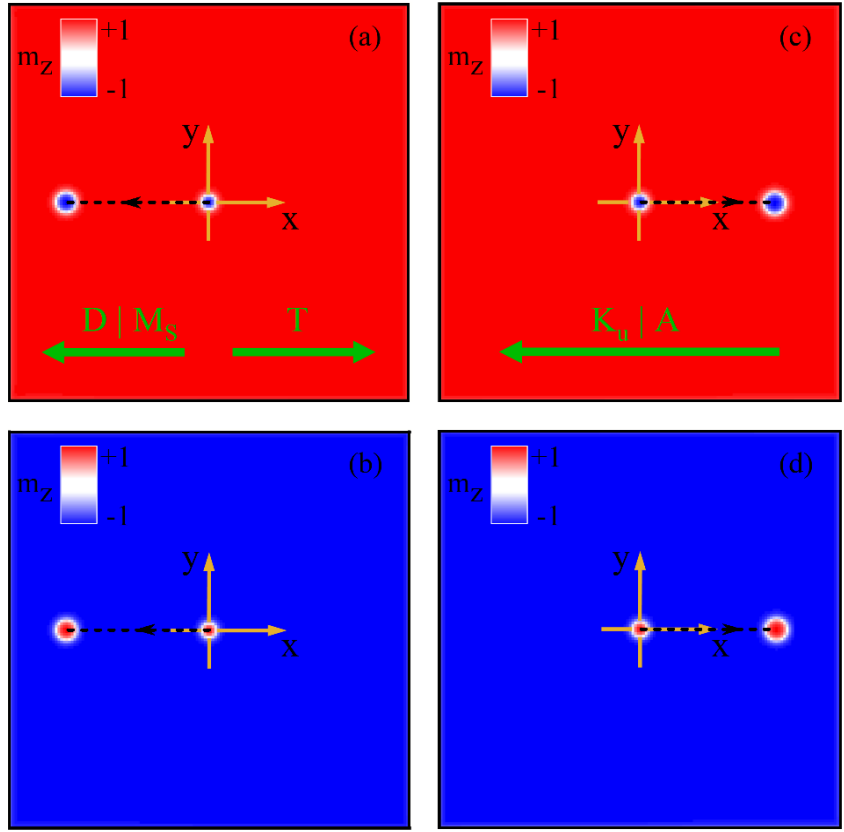

FIG. 5: (a) - (b) Skyrmion trajectory under a linear gradient of IDMI (or $M_S$ or thermal gradient), in the top layer and in the bottom layer of the SAF, respectively. (c) – (d) Skyrmion trajectory under a linear gradient of $K_u$ (or $A$), in top layer and in the bottom layer of the SAF, respectively.

III. SUMMARY AND CONLUSIONS

In summary, we analyzed the effect of linear thermal gradients on the skyrmion motion in single and multilayered thin films with interfacial DMI parameter in different systems through micromagnetic simulations and the Thiele formalism. We focused on the role of the entropic and magnonic torques. The former originates from the temperature dependence of the magnetic parameters, which scale with temperature. We observed opposite skyrmion motion directions in a single-layer FM and a multilayer. In the single-layer case, the skyrmion moves to the hotter region, whereas, in the multilayer case, the skyrmion moves towards the colder region, in qualitative agreement with experimental results [35]. We attributed this difference to the distinct scaling relations characterizing the two systems, and to the existence of a magnetostatic field gradient linked to the variation of $M_S$ which cannot be neglected in magnetic multilayers. We also showed that the magnonic torque promotes the skyrmion motion towards the hotter region with stochastic dispersion of the trajectories around averaged Hall angle. In a SAF, the skyrmion motion due to the entropic torque occurs with a zero Hall angle towards either the colder or the hotter region according to the scaling relations considered.

Our results show the importance of the design of proper temperature dependence of the magnetic parameters and can have a fundamental impact in the future development and design of skyrmion devices in the field of Skyrmion-Caloritronics.

ACKNOWLEDGMENT

This work was supported by the Project No. PRIN 2020LWPKH7 funded by the Italian Ministry of University and Research, and by the PETASPIN association (www.petaspin.com). JB acknowledges support from the Royal Society through a University Research Fellowship. E.R. acknowledges the economical support of the COST Action CA 17123 (MAGNETOFON) within the STSM program. We would like to acknowledge networking support from COST Action No. CA17123 "Ultrafast opto magneto electronics for nondissipative information technology."